\documentclass{mem}
\usepackage{natbib}\usepackage{txfonts}\usepackage{balance}
\usepackage{graphicx}
\begin{document}
\def\teff{$T\rm_{eff }$}
\def\kms{$\mathrm {km s}^{-1}$}

\newcommand{\linederiv}[2]{\mathrm{d}{#1}/\mathrm{d}{#2}}
\newcommand{\mathderiv}[2]{\frac{\mathrm{d}{#1}}{\mathrm{d}{#2}}}

\newcommand{\mathpartderiv}[2]{\frac{\partial{#1}}{\partial{#2}}}
\newcommand{\linepartderiv}[2]{{\partial{#1}}/{\partial{#2}}}


\title{
Overshoot Inwards from the Bottom of the Intershell Convective Zone in (S)AGB stars
}

   \subtitle{}
   
   \author{
     J. C. Lattanzio\inst{1},
     C. A. Tout\inst{2},
     E. V. Neumerzhitckii\inst{1},
     A. I. Karakas\inst{1},
     and
     P. Lesaffre\inst{3},
          }

\institute{
  Monash Centre for Astrophysics,
  School of Physics and Astronomy,
  Monash University, 3800, Australia
\and
Institute of Astronomy, The Observatories, 
Madingley Road,
Cambridge, CB3 0HA, UK
\and
Laboratoire de radioastronomie, LERMA,
Observatoire de Paris, École Normale Supérieure (UMR 8112 CNRS),
75231 Paris Cedex 05, France
}

\authorrunning{Lattanzio et al.}

\titlerunning{Intershell Overshoot in (S)AGB Stars}

\abstract{
  We estimate the extent of overshooting inwards from the bottom of the intershell convective
  zone in thermal pulses in (S)AGB stars. We find that the buoyancy
  is so strong that any overshooting should be negligible. The temperature inversion at the bottom of the convective zone
  adds to the stability of the region. Any mixing that occurs in this region is highly unlikely to be due to
  convective overshooting, and so must be due to another process.
\keywords{Stars: abundances --
Stars: AGB -- Stars: Evolution -- Stars: mixing
}
}
\maketitle{}

\section{Introduction}

Thermal pulses on the AGB and SAGB drive a convective zone that reaches from
the flashing helium shell almost to the convective
envelope. This is known as the Inter-Shell Convective Zone (ISCZ). The composition of this region is
mostly ashes from the hydrogen shell mixed
with the products of the rapid He burning that is driving the convection. Typically we have about $70$\% He and $25$\% C. This material is
later incorporated into the envelope by third dredge-up, and is
responsible for much of the evolution of the surface composition of AGB stars.

\section{Overshooting and its Consequences}
The Schwarzschild criterion for convection is commonly used to
determine the borders of convective regions. However it simply finds a radial
level where the buoyancy (the acceleration) is zero.
Any fluid moving
within the convective region toward the boundary reaches the boundary with a
finite velocity. Conservation of momentum
ensures that the fluid travels beyond
the Schwarzschild border, and this is the original definition of overshooting. 
Whilst the existence of overshooting is clear, it's \emph{extent\/} is another matter.

It is common to use the diffusion equation to 
calculate the mixing in convective regions. This is not strictly correct 
because mixing is \emph{advective\/} and not  \emph{diffusive\/}.
Nevertheless, this  provides a way to
calculate the composition in a region where the mixing is not sufficient to produce homogeneity. The mixing is determined by
the diffusion parameter $D$ that is used in the diffusion equation 

\begin{equation}
  {\mathderiv{X_i}{t}} = \left( \mathpartderiv{X_i}{t}\right)_{\rm nuc} +
                        \mathpartderiv{\ }{M_r}\left[ (4\pi r^2\rho)^2 D \mathpartderiv{X_i}{M_r} \right].
\end{equation}

A simple way to include overshooting is
to determine a (non-zero) $D$ outside the convective region. 
The pioneering work by \cite{heretal97} was motivated by multi-dimensional hydrodynamical calculations
of \cite{freetal96} which found a roughly exponential fall off in the velocity beyond the Schwarzschild boundary.
This was implemented by using $D=D_{\rm os}$, where
\begin{equation}
  D_{\rm os} = D_0 \exp \left(\frac{-2z}{H_{\rm V}}\right)
\end{equation}
and $D_0$ is the diffusion co-efficient near the Schwarzschild boundary.
Here $D_0 = v_0 H_{\rm P}$ and
the velocity scale-height $H_{\rm V}$ is assumed to be $H_{\rm V} = f H_{\rm P}$ where $f$ is a
constant.

This overshooting algorithm was applied to the ISCZ
in
AGB stars by \cite{heretal97}.
Mixing into the CO core alters the composition of the
ISCZ, increasing the C and O mass fractions. Indeed, \cite{heretal97} found that
the typical intershell mass fractions changed from He:C:O $\simeq$ 70:25:1 to 25:50:25.
This was found to be a better match to the observed abundances of
central stars of planetary nebulae and the PG1159 stars \citep{herwig00}.
But there were other changes, produced by the feedback of the mixing on the
stellar structure. First the extent of the dredge-up, as measured by the dredge-up
parameter $\lambda$, changed from about $0.7-0.8$ without overshooting to
values exceeding unity when overshooting was applied. This has a significant effect
on both the structural and chemical evolution of the star.
One very significant effect is seen in the production of  $s$-process elements.
In low- and intermediate mass stars we believe that
the neutron source is $^{13}$C, produced by partial mixing
of protons inward from the bottom of the convective envelope 
during third dredge-up events  \citep[for a recent review see][]{KL14} . At higher masses
the neutrons are thought to be provided by a different source:~CNO cycling
in the hydrogen shell produces $^{14}$N. When engulfed by the ISCZ at
the next pulse, two $\alpha$-captures on $^{14}$N produce $^{22}$Ne. At sufficiently high
temperatures, exceeding say 300 MK,
a further $\alpha$-capture can produce neutrons via $^{22}$Ne$(\alpha,$n)$^{25}$Mg. 
In the case of overshooting inwards from the ISCZ, the convective region 
is extended into the hot 
core with the result that the ISCZ experiences burning at higher temperatures.
This was shown by \cite{lugetal03} to produce significantly different abundances of $s$-process elements.
The
higher neutron densities also affect some branching points in the $s$-process
path with the result that some ratios no longer seem to match the observations
\citep[see][for details]{lugetal03}.

\begin{figure*}[h!]
\begin{center}
\resizebox{0.5\hsize}{!}{\includegraphics[clip=true]{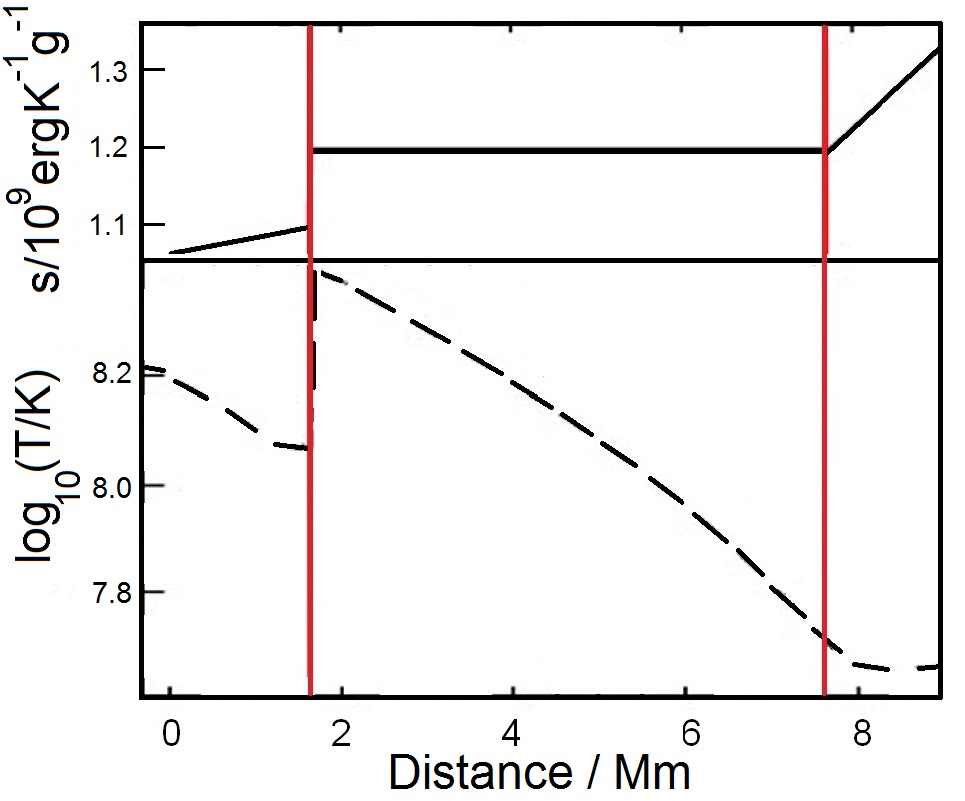}}
\caption{\footnotesize
Structure of the ISCZ during a thermal pulse. The top panel shows the entropy and the bottom panel shows the temperature distribution. The red lines show the Schwarzschild borders of convection. This plot is based on figure~5 of \cite{heretal06} for a $2M_\odot$ model with $Z=0.01$ and $f=0.016$.
}
\end{center}
\label{fig1}
\end{figure*}
\begin{figure*}[h!]
\begin{center}
\resizebox{0.7\hsize}{!}{\includegraphics[clip=true]{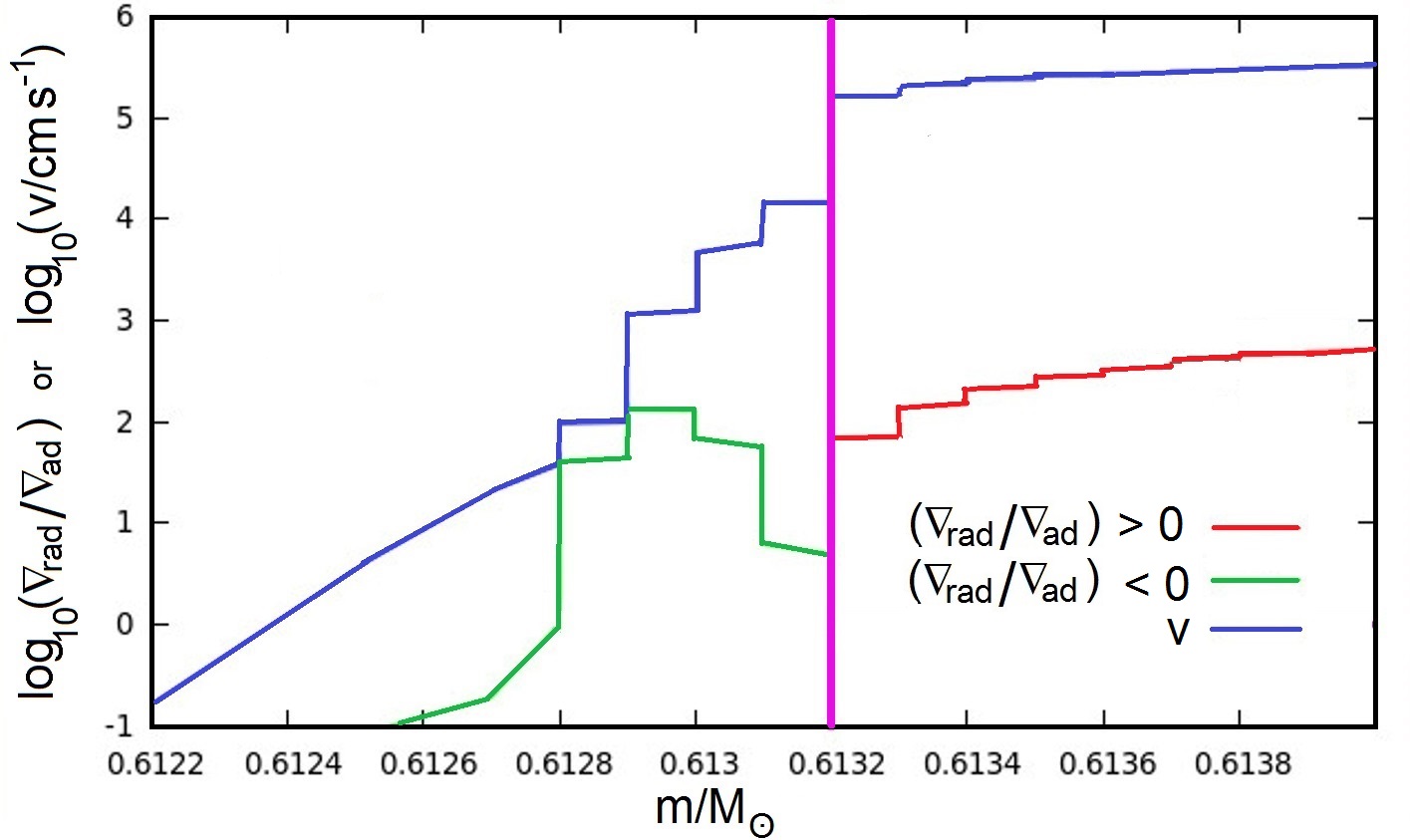}}
\caption{\footnotesize
Structure at the bottom of the ISCZ near the maximum strength of the 13th pulse for the model described in the text. The blue line shows log of the velocity while the red and green lines show log of the magnitude of $\nabla_{\rm rad}/\nabla_{\rm ad}$, coloured red and green where the ratio  is positive and negative respectively. Thus the green line shows the region of the temperature inversion. The magenta line shows the Schwarzschild boundary. 
}
\end{center}
\label{eta}
\end{figure*}

\section{Physics of the Region}
Thermal pulses drive the ISCZ, and the rapid injection of energy raises
the temperature locally. This produces a discontinuity in the temperature (and
so also the density) at the bottom of the ISCZ and also a temperature inversion. This is shown in Fig.~1, based on figure~5 of \cite{heretal06} who 
investigated the hydrodynamics of the ISCZ in 2D (and in one low resolution 3D 
calculation). They found that mixing outside the formal convective region was 
``orders of magnitude less efficient'' and showed an ``exponential decay'' in
moving away from the ISCZ. The calculations were not able to follow the evolution
for very long, but did find that ``mixing
of material into the convection zone from below $\ldots$ shows a significant upward trend at late times (after
several convective turnover times).''

We note that the temperature inversion means that the radiative temperature gradient
is negative. When subjected to a 
standard Schwarzschild analysis we find that the region is unconditionally stable 
because the 
temperature decreases when the density increases. Of course this is already known,  because ``outside the unstable region the material is stable" (Denissenkov, private communication). The question is one of how efficient is the mixing at the bottom of the ISCZ. 

We have calculated the evolution of a $3M_\odot$ model of solar metallicity using the
\textsc{Monstar} evolution code \citep[e.g.][]{GilPetal13}. We calculated two cases, one with no overshooting and 
one with the \cite{heretal97} prescription with $f=0.02$. Fig.~2 shows the bottom of the ISCZ near the maximum of the 13$^{\rm th}$ pulse. Note that the overshoot algorithm produces a significant velocity in the region of the temperature inversion. The buoyancy here is opposed to the penetration of the convective motion, but the buoyancy is not included in the algorithm which simply assigns a velocity. Does a naive application of the overshooting formula produce mixing that is consistent with the structure? We now try to estimate the strength of the restoring force that opposes the mixing enforced by the algorithm.

\section{Estimates of the overshoot}

We make some basic estimates of the extent of the overshooting inward from the
bottom of the ISCZ. Consider an eddy of volume $V_{\rm e}(r)$ and density $\rho_{\rm e}(r)$ at the bottom of the ISCZ.
Let the background density of the star be $\rho(r)$. Hence the eddy will feel a
restoring buoyancy force
$F_{\rm b} = g\,\Delta m $
where 
$\Delta m = \left( \rho(r) - \rho_{\rm e}(r)  \right) V_{\rm e}$
and the equation of motion is
$$F = \rho_{\rm e} V_{\rm e} g = g\,\Delta m $$ so that
$$a_{\rm e} = \frac{{\rm d}}{{\rm d}r} \left( \frac{1}{2}v^2 \right) = g(r)\, \left(\frac{\rho(r)-\rho_{\rm e}(r)}{\rho_{\rm e}(r)}\right).$$
Integrating from the bottom of the ISCZ, where $r = r_0$ and $v=v_0$ to the point $r_1$ where $v_1=0$ we find
$$\int\limits_{v_0}^{v_1} {\rm d}\left( \frac{1}{2} v^2  \right) = \frac{1}{2} v_0^2=
\int\limits_{r_0}^{r_1} g(r)\,\left[ \frac{\rho(r) - \rho_{\rm e}(r)}{\rho_{\rm e}(r)}  \right] {\rm d}r.$$ 
We make the usual assumption that the eddy moves adiabatically. Our procedure is to integrate the RHS downward until we reach a position $r_1$ where the integral is equal to $\frac{1}{2}\, v_0^2$. The depth $r_1$ is then the radial extent of the overshooting before buoyancy stops the eddy. We ignore the energy required to move material out of the path of the eddy, so our estimate for $r_1$ is a maximum. Further, for the velocity $v_0$ at the bottom of the ISCZ we take the largest velocity in the ISCZ, so that again our $r_1$ is a maximum.

\begin{table}[]
\centering
\caption{Estimates of the extent of overshooting}
\label{my-label}
\begin{tabular}{cccc}
 &{\bf Model}  &{\bf Phase}  &{\bf Distance (m)} \\
 &150   &Start of pulse    &1  \\
 &170   &Middle of pulse   &6  \\
 &1200  &Maximum of pulse  &200  \\
\end{tabular}
\end{table}

We performed this procedure at three times during the
13$^{\rm th}$ pulse. The results are shown in Table~1. In each case the extent of the overshooting is less than one mesh point in the 1D model. Of course, if we were to resolve this region (and the tiny change of composition) it may cause some feedback such that these
estimates are inaccurate. There are also many approximations and idealizations in the calculation. Nevertheless, it appears that the eddy faces a very strong
restoring force below the ISCZ, largely due to the temperature inversion. It seems that the eddy finds it very difficult to mix far enough to change
the composition in the way that seems favoured by comparisons with the composition of PG1159 stars for example.

\section{Conclusion}
Mixing within stars continues to be a problem. How to handle the expected overshooting, entrainment and other mixing processes at convective borders is a particularly pressing problem. It affects most phases of evolution. Together we call these processes convective border mixing. It is perhaps a more useful term than overshooting which may be just one of the processes taking place.

We have investigated the effect of buoyant overshooting at the bottom of the ISCZ during thermal pulses in AGB stars. Our idealized 1D analytic theory argues 
for a small to negligible extent of 
mixing owing to \emph{overshooting\/}. 
We note that models which enforce mixing, despite the arguments presented in this paper, seem to match the observations better. How do we understand this? While we feel rather safe in concluding that momentum-based \emph{overshooting\/} is unlikely to produce substantial mixing, we cannot rule out some other mixing  process, such as gravity waves of shear instabilities, as being more effective. Perhaps the evidence is indicating that another process is 
responsible for the required mixing. In any case, this does not seem to be classical overshooting.

\begin{acknowledgements}
  We thank Alessandro Chieffi, David Arnett and Peter Hoeflich for discussions. CAT thanks Churchill College for his fellowship.
\end{acknowledgements}

\bibliographystyle{aa}

\begin{thebibliography}{}

\bibitem[{Freytag et al.(1996)}]{freetal96}
  Freytag, B.,  et al.~1996, A\&A, 313, 497

  
\bibitem[Gil-Pons et al.(2013)]{GilPetal13}
Gil-Pons, P., et al.,~2013, A\&A, 557, 106

\bibitem[{Herwig et al.~(1997)}]{heretal97}
  Herwig, F., et al.~1997, A\&A, 324, L81


\bibitem[{Herwig(2000)}]{herwig00} 
  Herwig, F., 2000, A\&A, 360, 952

\bibitem[{Herwig et al.(2006)}]{heretal06}
  Herwig, F.,  et al.~2006, ApJ, 642, 1057

\bibitem[{Karakas and Lattanzio~(2014)}]{KL14}
  Karakas, A. I., and Lattanzio, J. C.,~2014, PASA, 31, 30


\bibitem[{Lugaro et al.(2003)}]{lugetal03} 
  Lugaro, M., et al.~2003, ApJ, 586, 1305




\end{thebibliography}

\end{document}